\documentclass{ws-procs9x6}
\def\trimmarks{}
\newcommand{\etal}{et al.}


\def\msun{\,{\rm M_\odot}}

\def\spose#1{\hbox to 0pt{#1\hss}}
\def\lta{\mathrel{\spose{\lower 3pt\hbox{$\mathchar"218$}}
\raise 2.0pt\hbox{$\mathchar"13C$}}}
\def\gta{\mathrel{\spose{\lower 3pt\hbox{$\mathchar"218$}}
\raise 2.0pt\hbox{$\mathchar"13E$}}}

\def\HI{\hbox{H~$\scriptstyle\rm I\ $}}
\def\HII{\hbox{H~$\scriptstyle\rm II\ $}}

\def\HeII{\hbox{He~$\scriptstyle\rm II\ $}}
\def\HeIII{\hbox{He~$\scriptstyle\rm III\ $}}

\def\arg#1{{\it#1\/}}

\begin{document}

\title{The Dawn of Galaxies\footnote{\uppercase{T}o appear in the proceedings
of the \uppercase{XXI} \uppercase{T}exas \uppercase{S}ymposium on 
\uppercase{R}elativistic 
\uppercase{A}strophysics, \uppercase{D}ecember 9--13 2002, 
\uppercase{F}lorence, \uppercase{I}taly.}}

\author{P. MADAU and M. KUHLEN}

\address{Department of Astronomy and Astrophysics, University of California,\\
1156 High Street, Santa Cruz, CA 95064, USA}

\maketitle

\abstracts{
The development of primordial inhomogeneities into the non-linear regime
and the formation of the first astrophysical objects within dark matter halos 
mark the transition from
a simple, neutral, cooling universe -- described by just a few parameters -- 
to a messy ionized one -- the realm of radiative, hydrodynamic, and star
formation processes. The recent measurement by the {\it WMAP} satellite
of a large optical depth to electron scattering implies that this
transition must have begun very early, and that the universe was
reionized at redshift $z_{\rm ion}=17\pm 5$. It is an early generation of 
extremely metal-poor massive stars and/or `seed' accreting black holes 
in subgalactic halos that may have generated the ultraviolet 
radiation and mechanical energy that reheated 
and reionized most of the hydrogen in
the cosmos. The detailed thermal, ionization, and chemical enrichment history 
of the universe during the crucial formative stages around $z=10-20$ depends 
on the power-spectrum
of density fluctuations on small scales, the stellar initial mass function 
and star formation efficiency, a complex network of poorly
understood `feedback' mechanisms, and remains one of the crucial missing links 
in galaxy formation and evolution studies.  
}

\section{Introduction}

The last decade has witnessed great advances in our understanding of the 
high redshift universe. The pace of observational cosmology
and extragalactic astronomy has never been faster, and progress has been equally 
significant on the theoretical side. The key idea of currently popular 
cosmological scenarios, that primordial density 
fluctuations grow by gravitational instability driven by cold, collisionless
dark 
matter (CDM), has been elaborated upon and explored in detail through 
large-scale numerical simulations on supercomputers, leading to a 
hierarchical (`bottom-up') scenario of structure formation. In this model, 
the first objects to form are on subgalactic scales, and merge to 
make progressively bigger structures (`hierarchical clustering'). 
Ordinary matter in the universe follows the dynamics dictated by the 
dark matter until radiative, hydrodynamic, and star formation processes 
take over. Perhaps the most remarkable success of this theory has been 
the prediction of anisotropies in the temperature of the cosmic microwave
background (CMB) radiation at about the level subsequently measured 
by the {\it COBE} satellite and most recently by the {\it BOOMERANG}, 
{\it MAXIMA}, {\it DASI}, {\it CBI}, {\it Archeops}, and {\it WMAP} 
experiments.

In spite of some significant achievements in our understanding of the 
formation of cosmic structures, there are still many challenges facing 
hierarchical clustering theories, and many fundamental questions remain, 
at best, only partially answered. While quite successful in matching the
observed large-scale density distribution (like, e.g., the properties 
of galaxy clusters, galaxy clustering, and the statistics of the Lyman-$\alpha$
forest), CDM simulations appear to 
produce halos that are too centrally concentrated compared to the mass 
distribution inferred from the rotation curves of (dark matter-dominated) 
dwarf galaxies, and to predict too many dark matter subhalos compared to 
the number of dwarf satellites observed within the Local 
Group.\cite{nfw,vdb,moore,klypin} Another perceived problem (possibly 
connected with the `missing satellites'\cite{bullock}) is our inability 
to predict when, how, and to what temperature the universe was reheated and 
reionized, i.e. to understand the initial conditions of the galaxy formation
process.
While N-body$+$hydrodynamical simulations have convincingly shown that the 
intergalactic medium (IGM) -- the main repository of baryons at high redshift --
is expected to fragment into structures at early times in
CDM cosmogonies, the same simulations are much less able to predict the
efficiency with which the first gravitationally collapsed objects lit up the
universe at the end of the `dark ages'. The crucial processes 
of star formation, preheating and feedback (e.g. the effect of the heat 
input from the first generation of sources on later ones), and assembly of 
massive black holes in the nuclei of galaxies are poorly understood.\cite{loeb}
We know that at least some galaxies and 
quasars were already shining when the universe was less than $10^9$ yr old. But
when did the first luminous objects form, what was their nature, and what 
impact did they have on their environment and on the formation
of more massive galaxies? While the excess 
\HI absorption measured in the spectra of $z\sim 6$ quasars in the Sloan
Digital Sky Survey (SDSS) has been interpreted as the signature of the 
trailing edge of the cosmic reionization epoch\cite{becker,fan,george},
the recent detection by the {\it Wilkinson Microwave Anisotropy Probe} ({\it 
WMAP}) of a large optical depth to Thomson scattering, $\tau_e=0.17\pm 0.04$ 
suggests that the universe was reionized at higher redshifts, $z_{\rm ion}=
17\pm 5$.\cite{kogut,spergel} This is of course an indication of significant 
star-formation activity at very early times.

In this talk I will summarize some recent developments in our understanding 
of the dawn of galaxies and the impact that some of the earliest cosmic
structure may have had on the baryonic universe. 

\section{The dark ages}

In the era of precision cosmology we know that, at a redshift 
$z_{\rm dec}=1088\pm 1$, exactly $t_{\rm dec}=(372\pm 14) \times 10^3$ 
years after the big bang, the universe became 
optically thin to Thomson scattering\cite{spergel}, and entered a 
`dark age'.\cite{rees} At this epoch the electron fraction dropped below 
13\% (Figure 1), and the primordial radiation cooled below 3000 K, shifting 
first into the infrared and then into the radio. 
\begin{figure}[b]
\begin{center}
\vspace{-0.4cm}
\includegraphics[width=.49\textwidth]{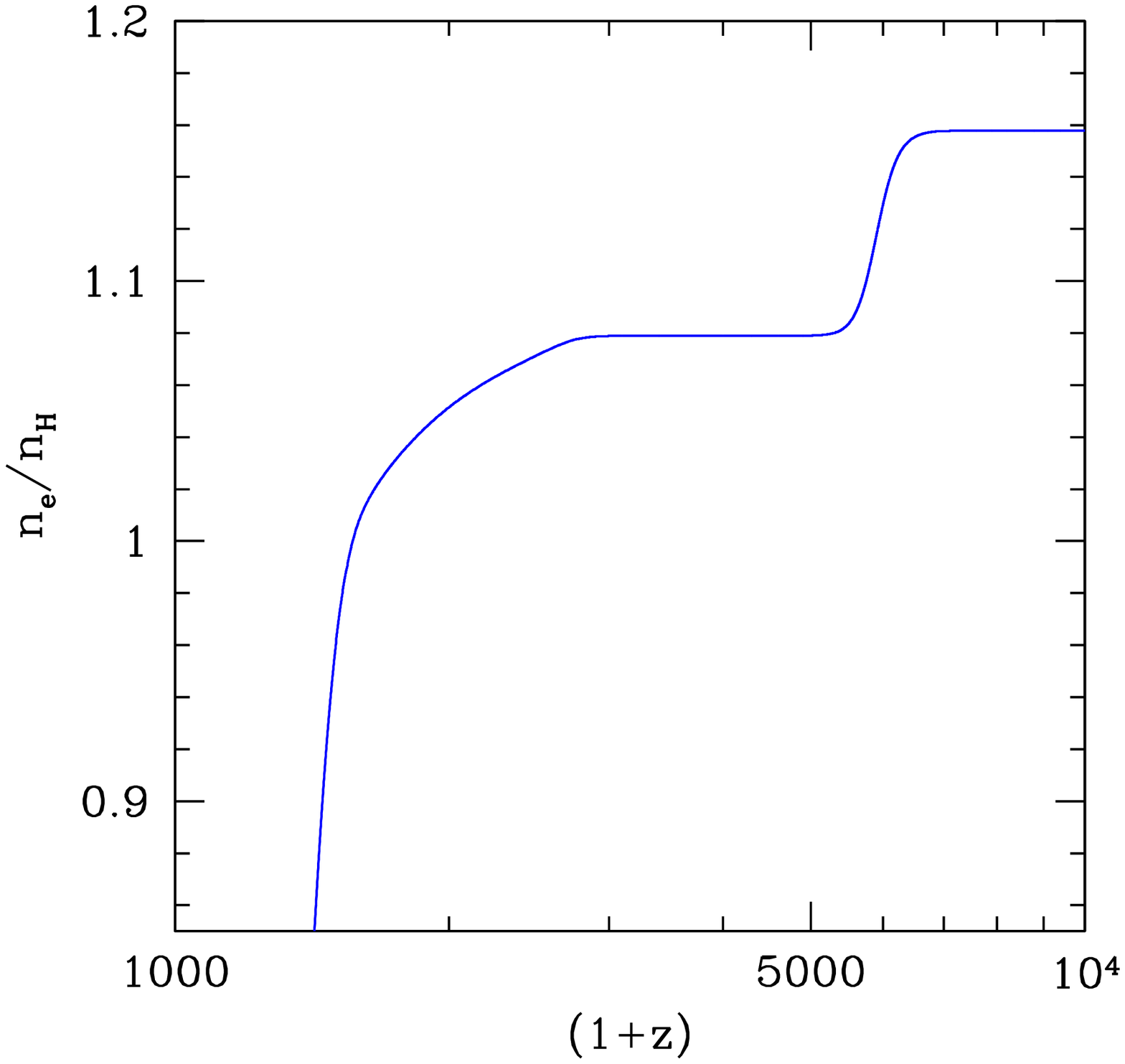}
\includegraphics[width=.49\textwidth]{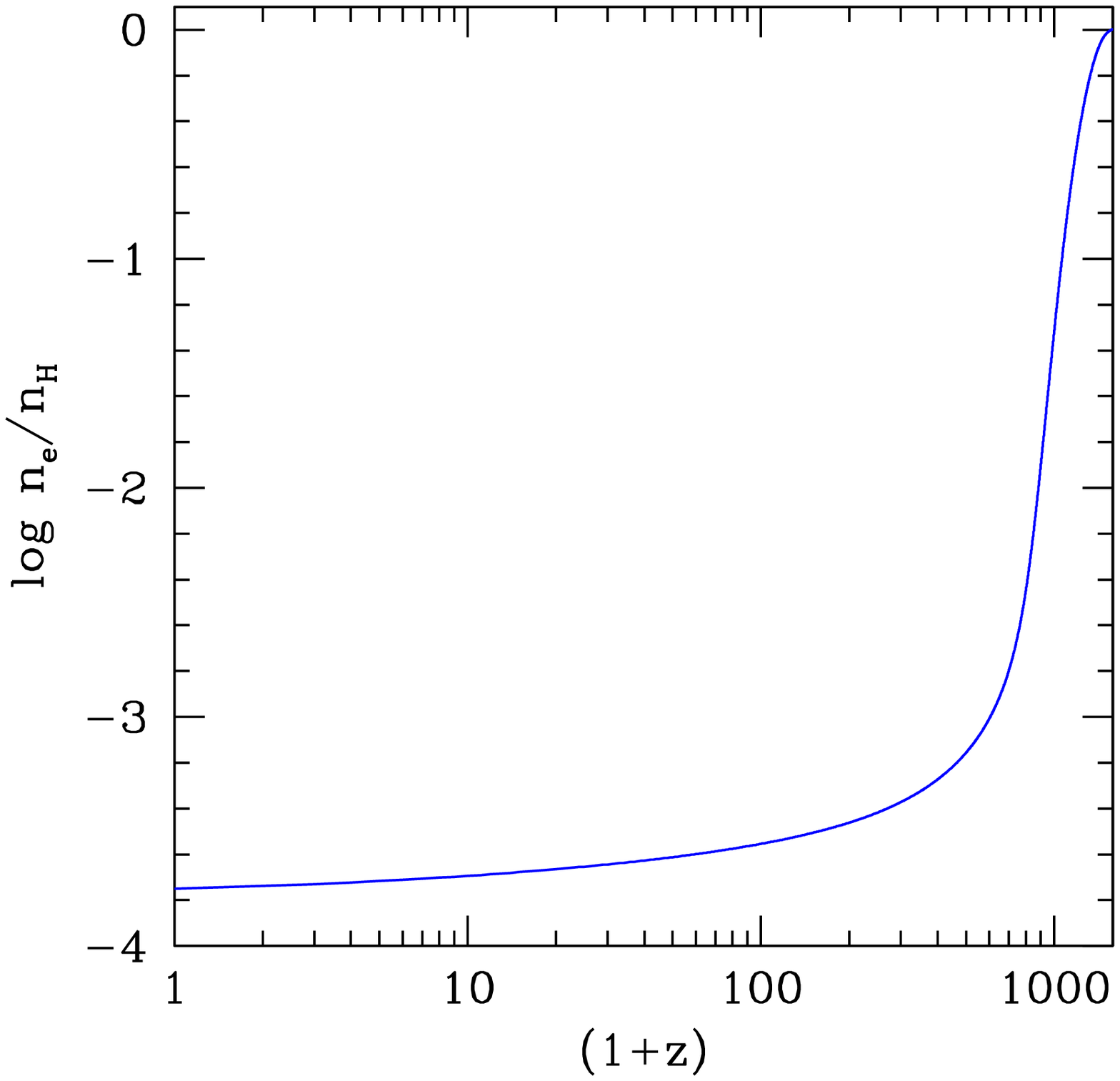}
\end{center}
\vspace{0.3cm}
\caption[]{\footnotesize Helium and hydrogen recombination for the {\it WMAP} 
parameters ($\Omega_M, \Omega_\Lambda, \Omega_b, h)=(0.29, 0.71, 0.045, 
0.7)$.\cite{spergel}
The step at earlier times in the left panel is due to the 
recombination of \HeIII into \HeII. We used the code RECFAST\cite{seager}
to compute the electron fraction. 
}
\end{figure}
We understand the microphysics of the
post-recombination universe well. The fractional ionization froze out
to the value $\sim 10^{-4.8}\Omega_M/(h\Omega_b)$: these residual
electrons were enough to keep the matter in thermal equilibrium with the
radiation via Compton scattering until a thermalization redshift
$z_t\simeq 800 (\Omega_bh^2)^{2/5}\simeq 150$, i.e. well after the universe
became transparent.\cite{peebles}
Thereafter, the matter temperature decreased as $(1+z)^2$ due to adiabatic
expansion (Figure 2) until primordial inhomogeneities in the density field
evolved into the non-linear regime.
\begin{figure}[b]
\begin{center}
\vspace{-0.4cm}
\includegraphics[width=.49\textwidth]{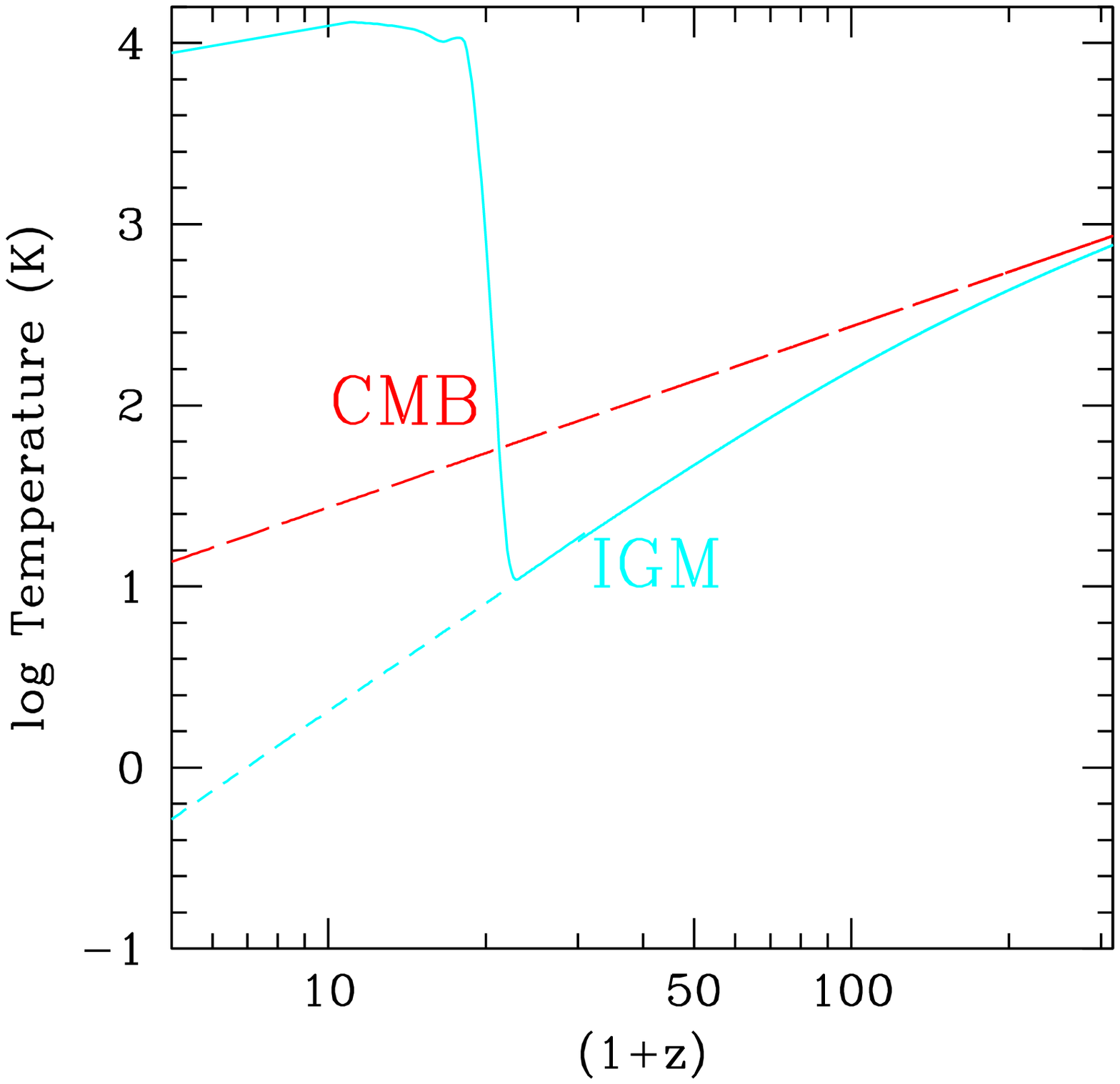}
\includegraphics[width=.49\textwidth]{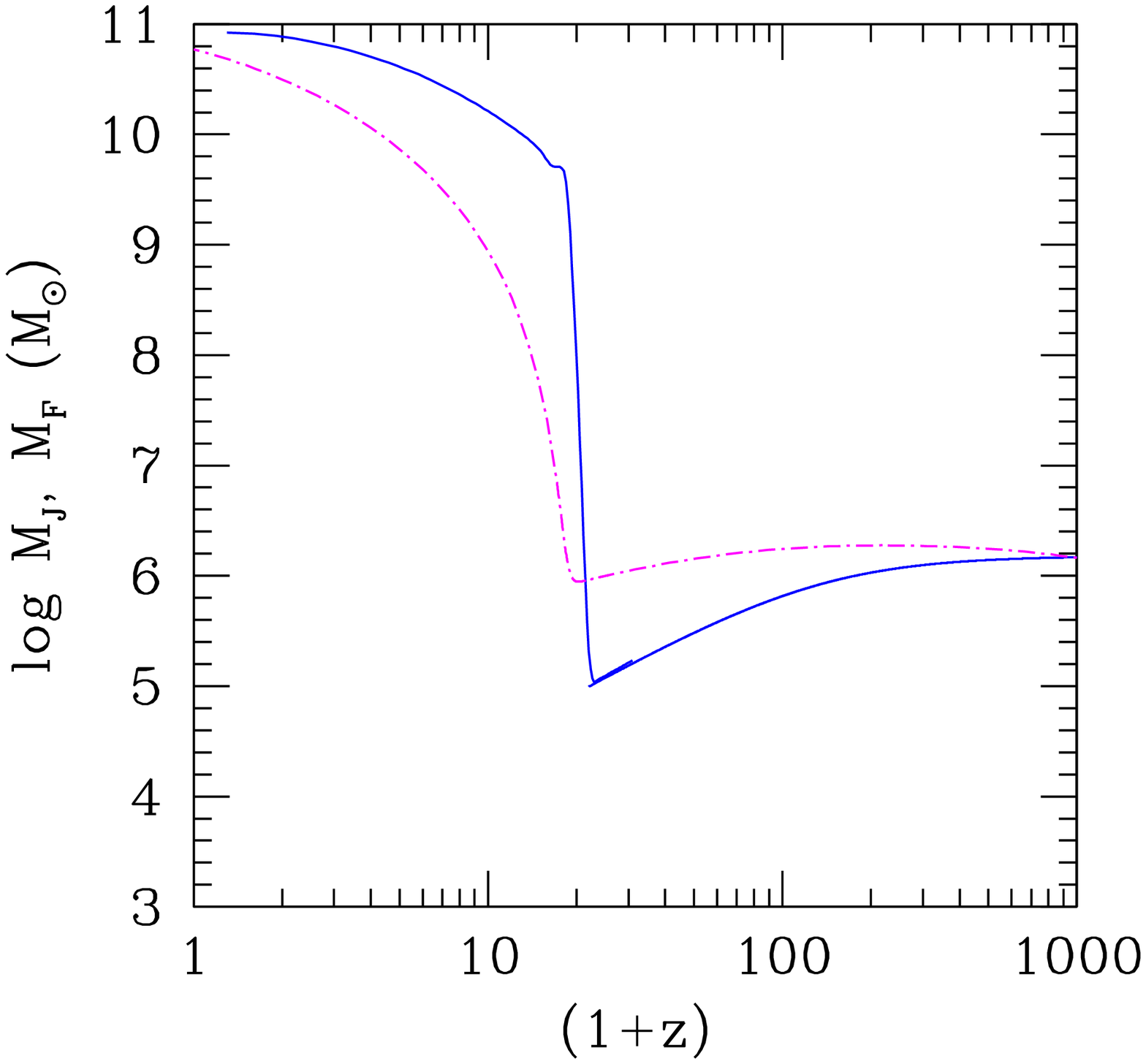}
\end{center}
\vspace{0.3cm}
\caption[]{\footnotesize {\it Left:} Evolution of the radiation
({\it long-dashed line}, labeled CMB) and gas ({\it solid line}, labeled
IGM) temperatures after recombination. The universe is assumed to be reionized
by ultraviolet radiation at $z\simeq 20$. The {\it short-dashed line} is the
extrapolated gas temperature in the absence of any reheating mechanism.
{\it Right:} Cosmological (gas $+$ dark matter) Jeans ({\it solid line}) 
and filtering ({\it dot-dashed line}) mass.
}
\end{figure}
The minimum mass scale for the gravitational aggregation of
cold dark matter particles is negligibly small. One of the
most popular CDM candidates is the neutralino: in neutralino CDM, collisional
damping and free streaming smear out all power of primordial density
inhomogeneities only below $\sim 10^{-7}\,\msun$.\cite{hofman}
Baryons, however, respond to pressure gradients and do not fall into dark
matter clumps below the cosmological Jeans mass (in linear
theory this is the minimum mass-scale of a perturbation where gravity
overcomes pressure),
\begin{equation}
M_J={4\pi \bar\rho\over 3}\left({5\pi k_BT\over 3G\bar\rho m_p\mu}\right)^{3/2}
\approx 2.5\times 10^5\,h^{-1}\,\msun (aT/\mu)^{3/2}\Omega_M^{-1/2}.
\end{equation}
Here $a=(1+z)^{-1}$ is the scale factor, $\bar \rho$ the total mass
density including dark matter, $\mu$ the mean molecular weight, 
and $T$ the gas temperature. The evolution of $M_J$ is shown in Figure 2. In
the post-recombination universe, the baryon-electron gas is thermally coupled
to the CMB, $T\propto a^{-1}$, and the Jeans mass is independent of redshift
and comparable to the mass of globular clusters, $M_J\approx 10^6\,\msun$. 
For $z<z_t$, the temperature of the baryons drops as $T
\propto a^{-2}$, and the Jeans mass decreases with time, $M_J\propto a^{-3/2}$.
This trend is reversed by the reheating of the IGM. The energy released
by the first collapsed objects drives the Jeans mass up to galaxy scales
(Figure 2): previously growing density perturbations decay as their mass drops
below the new Jeans mass. In particular, photoionization by the ultraviolet 
radiation from the first stars and quasars would heat the IGM to temperatures
of $\approx 10^4\,$K (corresponding to a Jeans mass 
$M_J\lta 10^{10}\,\msun$ at $z\simeq 20$),
suppressing gas infall into low mass halos and preventing new (dwarf)
galaxies from forming.

\section{Linear theory}

When the Jeans mass itself varies with time, linear gas fluctuations tend to
be smoothed on a (filtering) scale that depends on the full thermal
history of the gas instead of the instantaneous value of the sound 
speed.\cite{gnhui} From linear perturbation analysis, and for a flat 
universe at high redshift, the growth of density fluctuations in the gas 
is suppressed for comoving wavenumbers $k>k_F$, where the filtering scale 
$k_F$ is related to the Jeans wavenumber $k_J$ by\cite{gnedin}
\begin{equation}
{1 \over k^2_F(a)} = {3 \over a} \int_0^a {da^\prime\over k_J^2(a')}
[1-(a^\prime/a)^{1/2}].
\end{equation}
Here $k_J\equiv (a/c_s)\sqrt{4\pi G \bar{\rho}}$, and $c_s$ is the 
sound speed. This expression for $k_F$ accounts for an arbitrary thermal
evolution of the IGM through $k_J(a)$. Corresponding to the
critical wavenumber $k_F$ there is a critical (filtering) mass $M_F$, defined 
as the mass enclosed in the sphere with comoving radius equal to $k_F$, 
\begin{equation}
M_F=(4 \pi /3) \bar{\rho} (2 \pi a/k_F)^3.
\end{equation}
The Jeans mass $M_J$ is defined analogously in terms of $k_J$. It is 
the filtering mass that is central to calculations of the effects 
of reheating and reionization on galaxy formation. The filtering 
mass for a toy model with early photoionization is shown in Figure 2: after 
reheating, the filtering scale is actually smaller than the Jeans scale. 
Numerical simulations of cosmological reionization confirm that 
the characteristic suppression mass is typically lower than the 
linear-theory Jeans mass.\cite{gnedin}

\section{The emergence of cosmic structure}
\begin{figure}[b]
\begin{center}
\vspace{-0.4cm}
\includegraphics[width=.67\textwidth]{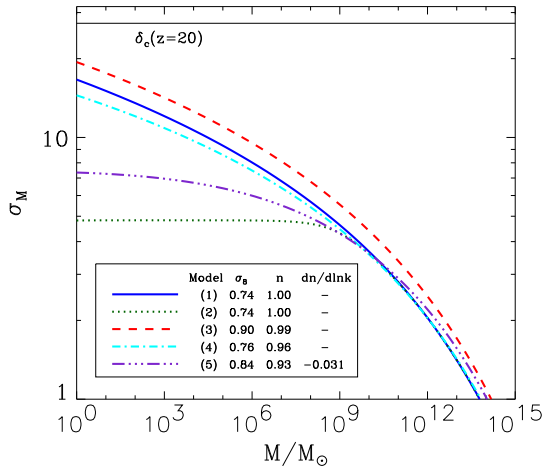}
\end{center}
\vspace{0.3cm}
\caption[]{\footnotesize The variance of the matter-density field vs. mass $M$, 
for different power spectra.
All models assume a `concordance' cosmology with parameters ($\Omega_M, 
\Omega_\Lambda, \Omega_b, h)=(0.29, 0.71, 0.045, 0.7)$.
{\it Solid curve:}  standard $\Lambda$CDM with no tilt, cluster normalized.
{\it Dotted curve:} $\Lambda$WDM with a particle mass $m_X=2\,$keV, cluster 
normalized, no tilt.
{\it Dashed curve:} tilted {\it WMAP} model, {\it WMAP} data only.
{\it Dash-dotted curve:} tilted {\it WMAP} model, including 2dFGRS and 
Lyman-$\alpha$ data.
{\it Dash-triple dotted curve:} running spectral index {\it WMAP} model,
including 2dFGRS and Lyman-$\alpha$ data. Here $n$ refers to the spectral 
index at $k=0.05\,\mbox{Mpc}^{-1}$. The horizontal line at the top of the
figure shows the value of the extrapolated collapse overdensity $\delta_c(z)$
at $z=20$.}
\end{figure}
As mentioned in the introduction, some shortcomings on galactic and sub-galactic
scales of the currently favored model of hierarchical galaxy formation in a
universe dominated by CDM have recently appeared. The significance of 
these discrepancies is still debated, and `gastrophysical' solutions 
involving feedback mechanisms may offer a possible way out. Other models
have attempted to solve the apparent small-scale problems of CDM at a more
fundamental level, i.e. by reducing small-scale power.
Although the `standard' $\Lambda$CDM model for structure formation assumes
a scale-invariant initial power spectrum of density fluctuations, 
$P(k)\propto k^n$ with $n=1$, the
recent {\it WMAP} data favor (but don't require) a slowly varying spectral 
index, $dn/d\ln k=-0.031^{+0.016}_{-0.018}$, i.e. a model in which 
the spectral index varies as a function of wavenumber $k$.\cite{spergel} 
This running 
spectral index model predicts a significanly lower amplitude of fluctuations 
on small 
scales than standard $\Lambda$CDM. The suppression of small-scale
power has the advantage of reducing the amount of substructure in galactic halos 
and makes small halos form later (when the universe was less dense) hence less
concentrated,\cite{nfw,zentner} relieving some of the problems of $\Lambda$CDM. 
But it makes early reionization a challenge.

Figure 3 shows the linearly extrapolated (to $z=0$) variance of 
the mass-density field smoothed on a scale of comoving radius $R$,
\begin{equation}
\sigma_M^2=\langle(\delta M/M)^2\rangle={1\over 2 \pi^2}\int_0^\infty
dk\, k^2 P(k) T^2(k) W^2(kR),
\end{equation}
for different power spectra. Here $M=H_0^2\Omega_MR^3/2G$ is the mass 
inside $R$, $T(k)$ is the transfer function for the matter density field (which 
accounts for all modifications of the primordial power-law spectrum due to 
the effects of pressure and dissipative processes), and $W(kR)$ is the 
Fourier transform of the spherical top-hat window
function, $W(x)=(3/x^2)(\sin x/x-\cos x)$. The value of the rms mass 
fluctuation in a 8 $h^{-1}\,$Mpc sphere, $\sigma_8\equiv
\sigma(z=0,R=8\,h^{-1}\,{\rm Mpc})$, has been fixed for the $n=1$ models
to $\sigma_8=0.74$, consistent with recent normalization by the $z=0$ X-ray 
cluster abundance constraint.\cite{reiprich} 

In the CDM paradigm, structure 
formation proceeds `bottom-up', i.e., the smallest objects collapse first, 
and subsequently merge together to form larger objects. It then follows 
that the loss of small-scale power modifies structure formation most severely 
at the highest redshifts, significantly reducing the number of 
self-gravitating objects then. This, of course, will make it more difficult 
to reionize the universe early enough.
\begin{figure}[b]
\begin{center}
\vspace{-0.5cm}
\includegraphics[width=.49\textwidth]{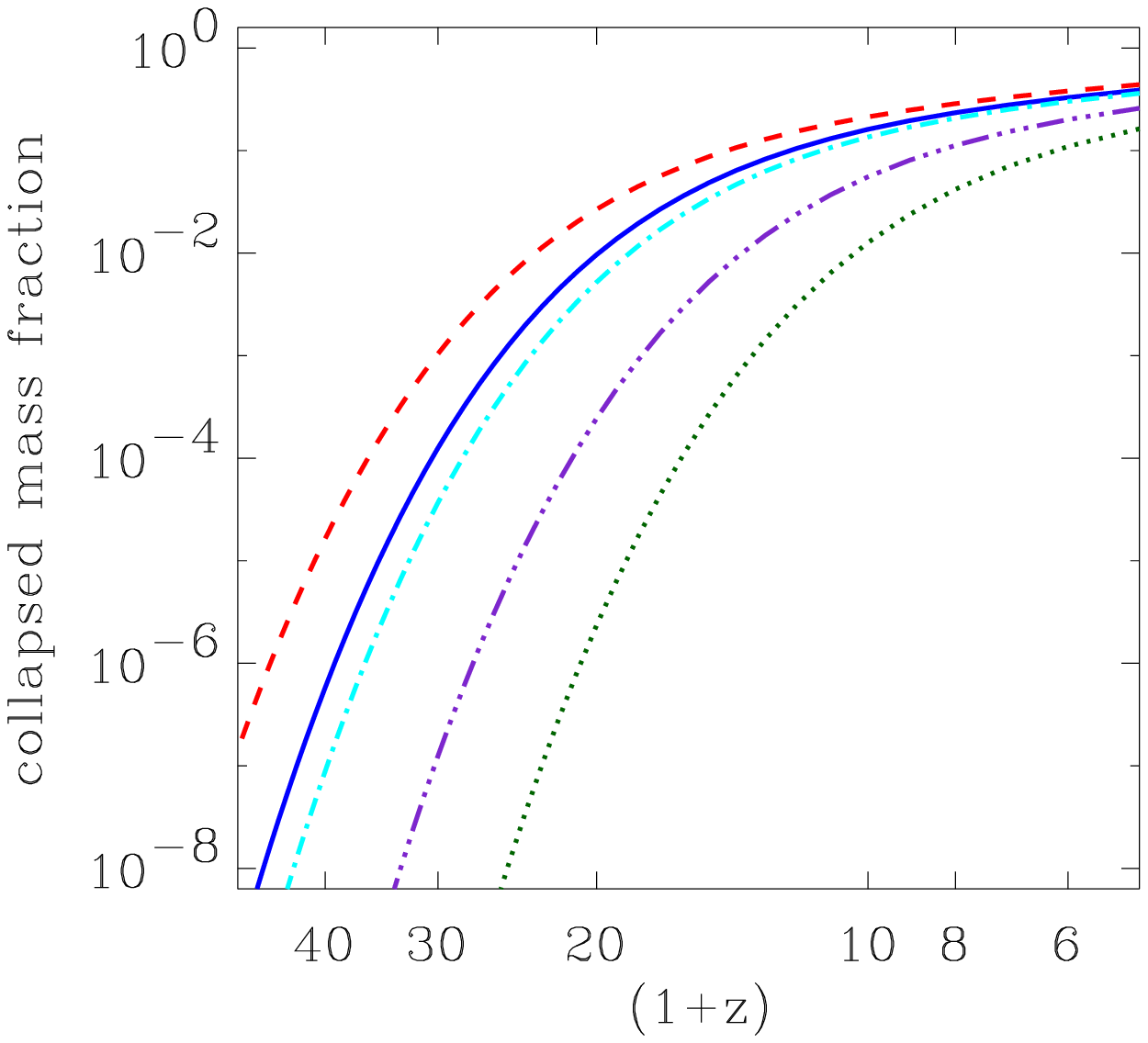}
\includegraphics[width=.49\textwidth]{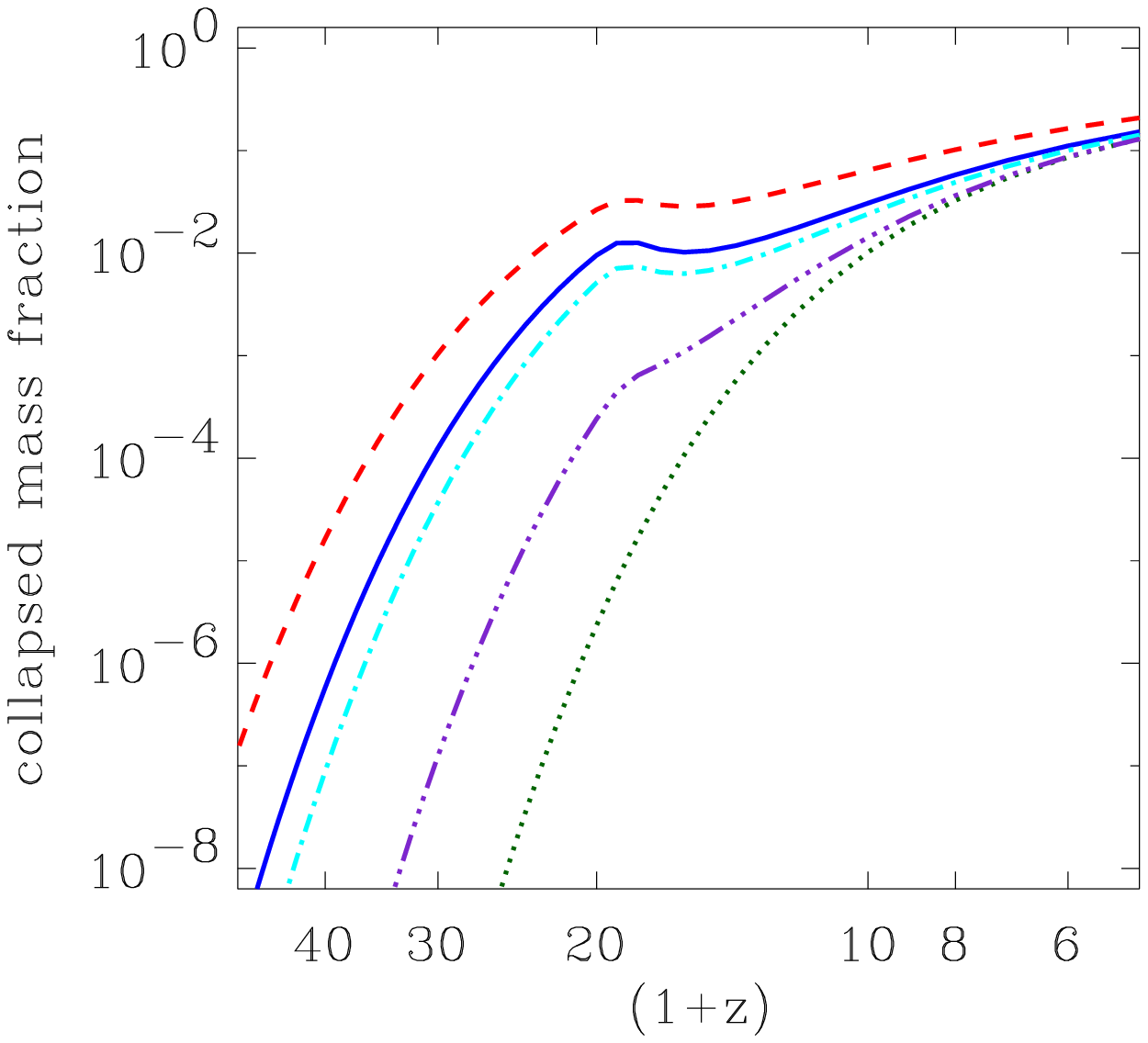}
\end{center}
\vspace{0.3cm}
\caption[]{\footnotesize Mass fraction in all collapsed halos above the 
filtering mass $M_F$ as a function of redshift, for different power spectra.
Curves are the same as in Figure 3. {\it Left panel:} filtering mass $M_F$ 
has been computed in the absence of reionization. {\it Right panel}: $M_F$ 
computed assuming the universe is reionized by ultraviolet radiation at 
$z\simeq 20$.   
}
\end{figure}
It has been argued,
for example, that one popular modification of the CDM paradigm, warm dark 
matter (WDM), has so little structure at high redshift that it is unable to 
explain the {\it WMAP} observations of an early epoch of 
reionization.\cite{spergel,barkana} And yet the {\it 
WMAP} running-index model may suffer from a similar problem.\cite{sbl} 
A look at Figure 3 shows that $10^6\,$M$_\odot$ halos 
will collapse at $z=20$ from $2.9\,\sigma$ fluctuations in a tilted 
$\Lambda$CDM model with $n=0.99$ and $\sigma_8=0.9$,
from $4.6\,\sigma$ fluctuations in a running-index model, and from  
$5.7\,\sigma$ fluctuations in a WDM cosmology. The problem is that scenarios with 
increasingly rarer halos at early times require even more extreme assumptions 
(i.e. higher star formation efficiencies and UV photon production rates) in 
order to be able to reionize the universe by $z\sim 17$ as favored by {\it 
WMAP}.\cite{sbl,wl,hh,cfw,cen} Figure 4 depicts the mass fraction in 
all collapsed halos with masses above the {\it filtering mass} for a case 
without reionization and one with reionization occuring at $z\simeq 20$. 
At early epochs this quantity appears to vary by orders of magnitude in 
different models! 

\section{The epoch of reionization}

Since hierarchical clustering theories provide a well-defined
framework in which the history of baryonic material can 
be tracked through cosmic time, probing the 
reionization epoch may then help constrain 
competing models for the formation of cosmic structures. 
Quite apart from uncertainties in the primordial power spectrum on small
scales, however, it is the astrophysics of baryons that makes us unable 
to predict 
when reionization actually occurred. Consider the following illustrative 
example:
  
Hydrogen photoionization requires more than one photon above 13.6 eV 
per hydrogen atom: of order $t/\bar t_{\rm rec}\sim 10$ (where 
$\bar t_{\rm rec}$ is the volume-averaged hydrogen recombination 
timescale) extra photons 
appear to be needed to keep the gas in overdense regions and filaments 
ionized against radiative recombinations.\cite{gnedin2,mhr}. 
A `typical' stellar population produces during its lifetime
about 4000 Lyman continuum (ionizing) photons per stellar proton.
A fraction $f\sim 0.25$\% of cosmic baryons must then
condense into stars to supply the requisite ultraviolet flux. This 
estimate assumes a standard (Salpeter) initial mass function (IMF), 
which determines the relative abundances of hot, high mass stars 
versus cold, low mass ones. 

The very first generation of stars 
(`Population III') must have formed, however, out of unmagnetized 
metal-free gas: numerical simulations of the fragmentation of pure
H and He molecular clouds\cite{bromm,abel} have shown that these 
characteristics likely led to a `top-heavy' IMF biased towards very 
massive 
stars (VMSs, i.e. stars a few hundred times more massive than the Sun), 
quite different from the present-day Galactic case. Metal-free VMSs
emit about $10^5$ Lyman continuum photons per stellar baryon\cite{bromm2},
approximately $25$ times more than a standard stellar 
population. A corresponding smaller fraction of cosmic baryons would have 
to collapse then into VMSs to reionize the universe, $f\sim 10^{-4}$.
There are of course further complications. Since, at zero 
metallicity, mass loss through radiatively-driven stellar winds is 
expected to be negligible\cite{kudri}, Population  III stars may 
actually die 
losing only a small fraction of their mass. If they retain their large 
mass until death, VMSs with masses $100\lta m\lta 250\,\msun$ will
encounter the electron-positron pair instability and disappear in a 
giant nuclear-powered explosion\cite{fryer}, leaving no compact 
remnants and polluting the universe with the first heavy 
elements. In still heavier stars, however, oxygen and
silicon burning is unable to drive an explosion, and complete collapse
to a black hole will occur instead.\cite{bond} 
Thin disk accretion onto a Schwarzschild black hole releases about 
50 MeV per baryon. The conversion of a trace amount of the total 
baryonic mass into early black holes, $f\sim 3\times 10^{-6}$, 
would then suffice to reionize the universe. 

\section{Preheating and galaxy formation}

Even if the IMF at early times were known, we still would remain uncertain about the fraction of cold gas 
that gets retained in protogalaxies after
the formation of the first stars (this quantity affects the global 
efficiency of star formation at these epochs) 
and whether -- in addition to ultraviolet radiation -- an early input of 
mechanical energy may also play a role in 
determining the thermal and ionization state of the IGM on large scales. 
The same massive stars that emit ultraviolet light also explode as 
supernovae 
(SNe), returning most of the metals to the interstellar medium of
pregalactic systems and injecting about $10^{51}\,$ergs per event 
in kinetic energy. A complex network of feedback 
mechanisms is likely at work in these systems, as the gas in shallow 
potential is more easily blown away,\cite{dekel} thereby quenching star 
formation. Furthermore, as the 
blastwaves produced by supernova explosions -- and possibly also by winds
from `miniquasars' -- sweep the surrounding intergalactic gas,
they may inhibit the formation of nearby low-mass 
galaxies due to `baryonic stripping'\cite{scanna}, and drive vast 
portions of the IGM to a significantly higher temperature than expected 
from photoionization,\cite{voit,madau,mfr,theuns,cb} so as to `choke off' the 
collapse of further galaxy-scale systems. Note that this type of 
global feedback is
fundamentally different from the `in situ' heat deposition commonly adopted in
galaxy formation models, in which hot gas is produced by supernovae
within the parent galaxy. We refer here to this
global early energy input as `preheating'.\cite{benson} Note that a large 
scale feedback
mechanism may also be operating in the intracluster medium: studies of
X-ray emitting gas in clusters show evidence for some form of
non-gravitational entropy input \cite{ponman}. The energy required there 
is at a level of $\sim 1\,$keV per particle, and must be injected either
in a more localized fashion or at late epochs in order not to violate
observational constraints on the temperature of the Lyman-$\alpha$ forest at
$z\sim 3$.
The thermal and ionization history of a preheated universe may be very 
different from one where hydrogen is photoionized. The gas will be heated up
to a higher adiabat, and collisions with hot electrons will be the dominant
ionization mechanism. The higher energies associated with 
preheating may doubly ionize helium at high-$z$, well before the `quasar 
epoch' at $z\sim 3$. Galaxy formation and evolution will also be
different, as preheating will drive the filtering mass above 
$10^{10}-10^{11}\,\msun$ and will tend to flatten the faint-end slope of the 
present-epoch galaxy luminosity function, in excellent agreement with the data
and without the need for SN feedback at late times.\cite{benson}

It is interesting to set some general constraints on the
early star-formation episode and stellar populations that may be
responsible for an early preheating of the IGM at the levels consistent 
with the temperature of intergalactic gas inferred at $z\approx 3$. Let us
characterize the energy input due to preheating by the energy per baryon,
$E_p$, deposited in the IGM at redshift $z_p$. We examine a homogenous 
energy deposition since the filling factor of pregalactic outflows is 
expected to be large.\cite{mfr,fl} Let $\Omega_*$ be the mass density of 
stars formed at $z_p$ in
units of the critical density, $E_{\rm SN}$ the mechanical energy
injected per SN event, and $f_w$ the fraction of that energy that is
eventually deposited into the IGM. Denoting with $\eta$ the number of
SN explosions per mass of stars formed, one can write 
\begin{equation}
{\Omega_* \over \Omega_b}={E_p\over f_w\eta E_{\rm SN} 
m_p}, 
\label{starb}        
\end{equation}
where $m_p$ is the proton mass. For a Salpeter IMF
between 0.1 and 100 M$_\odot$, the number of Type II SN explosions per 
mass of stars formed is $\eta=0.0074$ M$_\odot^{-1}$, assuming all stars 
above 8 M$_\odot$ result in SNe II. Numerical simulations of the dynamics of
SN-driven bubbles from subgalactic halos have shown that up to 40\% of the
available SN mechanical luminosity can be converted into kinetic energy of
the blown away material, $f_w\approx 0.4$, the remainder being radiated 
away.\cite{mori}~ With $E_{\rm SN}=1.2\times 10^{51}\,$ergs, 
equation (\ref{starb}) then implies 
\begin{equation}
\left({\Omega_*\over \Omega_b}\right)_{\rm sp}=0.05~(E_p/0.1~{\rm 
keV}).
\end{equation}
SN-driven pregalactic outflows efficiently carry metals into
intergalactic space.\cite{mfr} For a normal IMF, the total
amount of metals expelled in winds and final ejecta (in SNe or
planetary nebulae) is about 1\% of the input mass. Assuming a large
fraction, $f_Z=0.5$, of the metal-rich SN ejecta escape the shallow
potential wells of subgalactic systems, the star-formation episode
responsible for early preheating will enrich the IGM to a mean level
\begin{equation}
\langle Z\rangle_{\rm sp}={0.01\,\Omega_*\,f_Z\over \Omega_b}=
0.014\,Z_\odot~(E_p/0.1~{\rm keV}).
\end{equation}
The weak C IV absorption lines observed in the Lyman-$\alpha$ forest at
$z=3-3.5$ imply a minimum universal metallicity relative to solar in
the range $[-3.2]$ to $[-2.5]$.\cite{songaila}. Preheating energies in
excess of $0.1\,$keV appear then to require values of $\Omega_*$ and
$\langle Z\rangle$ that are too high, comparable to the total mass
fraction in stars seen today\cite{glazebrook} and in
excess of the enrichment of the IGM inferred at intermediate
redshifts, respectively.

The astrophysics of first light may not be as simple, however. The metal 
constraint
assumes that metals escaping from protogalaxies are evenly mixed into
the IGM and the Lyman-$\alpha$ clouds.\cite{voit} Inefficient mixing could
instead produce a large variance in intergalactic metallicities. The
metal abundances of the Lyman-$\alpha$ clouds may underestimate the average
metallicity of the IGM if there existed a significant warm-hot gas
phase component with a higher level of enrichment, as detected for
example in O VI.\cite{simcoe}  Today, the
metallicity of the IGM may be closer to $\sim 1/3$ of solar if the
metal productivity of galaxies within clusters is to be taken as
representative of the universe as a whole.\cite{renzini} 
Uncertainties in the early IMF make other preheating
scenarios possible and perhaps even more likely.  
Population III stars with main-sequence masses of 
approximately $140-260\,$ M$_\odot$ will encounter the electron-positron pair
instability and be completely disrupted by a giant nuclear-powered
explosion.\cite{heger} A fiducial 200 M$_\odot$ Population
III star will explode with a kinetic energy at infinity of $E_{\rm
SN}=4\times 10^{52}\,$ ergs, injecting about 90 M$_\odot$ of metals.
For a very `top-heavy' IMF with $\eta=0.005$
M$_\odot^{-1}$, equation (\ref{starb}) now yields
\begin{equation}
\left({\Omega_* \over \Omega_b}\right)_{\rm III}=
0.001~(E_p/0.1\,{\rm keV}),
\end{equation}
and a mean IGM metallicity 
\begin{equation}
\langle Z\rangle_{\rm III}={0.45\,\Omega_*\,f_Z\over \Omega_b}=
0.02\,Z_\odot~(E_p/0.1\,{\rm keV})
\end{equation}
(in both expressions above we have assumed $f_w=f_Z=1$). This scenario
can yield large preheating energies by converting only a small
fraction of the comic baryons into Population III stars, but tends to
produce too many metals for $E_p\gta 0.1\,$keV. The
metallicity constraint, of course, does not bound preheating from
winds produced by an early, numerous population of faint 
`miniquasars'.\footnote{Because the number density of {\it bright} 
quasi-stellar objects at $z>3$ is low\cite{fan01}, the thermal and kinetic
energy they expel into intergalactic space must be very large to have
a global effect, i.e. for their blastwaves to fill and preheat the
universe as a whole. The energy density needed for rare, luminous quasars to 
shock-heat the entire IGM would in this case violate the {\it COBE}
limit on $y$-distortion.\cite{voit}}~Accretion onto black holes releases 
50 MeV per baryon, and if a fraction
$f_w$ of this energy is used to drive an outflow and is ultimately
deposited into the IGM, the accretion of a trace amount of the total 
baryonic mass onto early black holes, 
\begin{equation}
{\Omega_{\rm BH}\over \Omega_b}={E_p\over f_w\,50\,{\rm MeV}}
=2\times 10^{-6}~f_w^{-1}\,(E_p/0.1\,{\rm keV}),
\end{equation}
may then suffice to preheat the whole universe. Note that this value
is about $50\,f_w$ times smaller than the density parameter of the 
supermassive variety found today in the nuclei of most nearby galaxies, 
$\Omega_{\rm SMBH}\approx 2\times 10^{-6}\,h^{-1}$.\cite{merritt}

\section{Conclusions}

The above discussion should make it clear that, despite much recent progress 
in our understanding of the formation of early cosmic structure and the 
high-redshift universe, the astrophysics of first light remains one of the 
missing links in galaxy formation and evolution studies. We are left very 
uncertain about the whole era from $10^8$ to $10^9$ yr -- the epoch of the
first galaxies, stars, supernovae, and massive black holes.
Some of the issues
discussed above are likely to remain a topic of lively controversy
until the launch of the {\it James Webb Space Telescope} ({\it JWST}), 
ideally suited
to image the earliest generation of stars in the universe.
If the first massive black holes form in pregalactic systems at very high
redshifts, they will be incorporated through a series of mergers into 
larger and larger halos, sink to the center owing to dynamical friction,
accrete a fraction of the gas in the merger remnant to become
supermassive, and form binary systems.\cite{volonteri} Their coalescence
would be signalled by the emission of low-frequency gravitational waves
detectable by the planned {\it Laser Interferometer Space Antenna} ({\it LISA}). 
An alternative way to probe the end of the dark ages and discriminate
between different reionization histories is through 21 cm tomography.\cite{mmr}
Prior to the epoch of full reionization, 21 cm spectral features will 
display angular structure as well as 
structure in redshift space due to inhomogeneities in the gas density field,
hydrogen ionized fraction, and spin temperature. Radio maps will show a
patchwork (both in angle and in frequency) of emission signals from \HI
zones modulated by \HII regions where no signal is detectable against the
CMB.\cite{cm} The search
at 21 cm for the epoch of first light has
become one of the main science drivers of the {\it LOw Frequency ARray}
({\it LOFAR}).  While remaining an extremely
challenging project due to foreground contamination from unresolved
extragalactic radio sources\cite{dimatteo} and free-free emission
from the same halos that reionize the universe\cite{om}, the
detection and imaging of large-scale structure prior to reionization
breakthrough remains a tantalizing possibility within range of the next
generation of radio arrays.

\section*{Acknowledgments}
We have benefited from many discussions with all our collaborators 
on the topics described here. PM acknowledges the hospitality of the Carnegie 
Observatories where part of this review was written. Support for this work was 
provided by NSF grant AST-0205738 and by NASA grant NAG5-11513.

\end{document}